\input harvmac.tex

\chardef\tempcat=\the\catcode`\@
\catcode`\@=11
\def\cyracc{\def\u##1{\if \i##1\accent"24 i%
    \else \accent"24 ##1\fi }}
\newfam\cyrfam
\font\tencyr=wncyr10
\def\cyr{\fam\cyrfam\tencyr\cyracc}
\def\MM{\overline{\cal M}}
\lref\amm{H.~Airault, H.~McKean, J.~Moser, ``Rational and elliptic
solutions of the KdV equation and a related many-body problem'', 
Comm. Pure. Appl. Math. {\bf 30} (1977) no. 1, 95--148}
\lref\krichphong{I.~Krichever, D.~H.~Phong, `` On 
the integrable geometry of soliton equations 
and N=2 supersymmetric gauge theories'', 
hep-th/9604199, J.~Diff.~Geom. {\bf 45} (1997) 349-389}
\lref\dijkgraaf{R.~Dijkgraaf,  
``Instanton Strings and HyperK\"ahler Geometry'',
hep-th/9810210}
\lref\sepex{E.~Sklyanin, ``Separation of variables in the classical
integrable ${\rm SL}(3)$ magnetic chain. Comm. Math. Phys. 150 (1992),
 no. 1, 181--191. }
\lref\sentwo{A.~Sen, ``$BPS$ states on a three brane probe'', hep-th/9608005}
\lref\vafaf{C.~Vafa, ``Evidence for $F$-theory'', hep-th/9602022}
\lref\schwarzm{J.~Schwarz, ``The Power of $M$-theory'', hep-th/9508143}
\lref\wittensol{E.~Witten,``Solution of four dimensional field
theories via $M$-theory'', hep-th/9703166}
\lref\naklect{H.~Nakajima ,"Lectures on Hilbert schemes of points on
surfaces", H.~Nakajima's homepage}
\lref\nakjack{H.~Nakajima,
``Jack polynomials and Hilbert schemes of points on surfaces'',
 alg-geom/9610021}
\lref\nakheis{H.~Nakajima,
``Heisenberg algebra and Hilbert schemes of points on projective
 surfaces'',
 alg-geom/9507012}
\lref\setkap{A.~Kapustin, S.~Sethi, 
``The Higgs Branch of Impurity Theories'', hep-th/9804027, Adv. Theor. Math. 
Phys 
{\bf 2} (1998) 571 - 591}
\lref\taylor{W.~Taylor, ``D-Brane Field Theory on a Compact Space',
hep-th/9611042, Phys. Lett. {\bf B}394 (1997) 283-287}
\lref\nakaff{H.~Nakajima,
``Instantons and affine Lie algebras'',
alg-geom/9510003}
\lref\Hurt{J.~Hurtubise, ``Integrable systems and algebraic surfaces",
Duke.~Math.~Journal {\bf 83} (1996) 19-50}
\lref\wilson{G.~Wilson, ``Collisions of Calogero-Moser particles and
an  adelic Grassmanian", Inv. Math. {\bf 133} (1998), no. 1,  1-41}
\lref\nhol{N.~Nekrasov, ``Holomorphic bundles and many-body
systems'', hep-th/9503157,
Comm. Math. Phys. {\bf 180}  (1996) 587}
\lref\hi{N.~Hitchin, ``Stable bundles and integrable systems'', Duke. Math. 
Journ.
{\bf 54}  (1987), 91-114}

\lref\oldsov{H.~Flaschka and D.~W.~McLaughlin, Progr. Theor. Phys. {\bf 55} 
(1976) 438-456\semi
I.~M.~Gelfand and L.~A.~Diki\~i, Func. Anal. Appl. {\bf 13} (1979) 8-20\semi
S.~P.~Novikov and A.~P.~Veselov Proc. Steklov Inst. Math. {\bf 3} (1985) 53-65}

\lref\hid{N.~Hitchin, ``The self-duality equations on a Riemann surface'',
Proc. London Math. Soc. {\bf 55} (1987) 59-126 }
\lref\Mu{S.~Mukai, " Symplectic structure of the moduli space
on a abelian or K3 surface", Invent.~Math. {\bf 83} (1984) 101-116 }
\lref\DonEinLaz{R.~Donagi, L.~Ein, R.~Lazarsfeld,
alg-geom/9504017;
``Nilpotent cones and sheaves on $K3$ surfaces '',
Birational algebraic geometry (Baltimore, MD, 1996), 51--61,
Contemp. Math., 207, Amer. Math. Soc., Providence, RI, 1997 }
\lref\Beaui{A.~Beauville, ``Syst\`emes Hamiltoniens compl\'etement
int\`egrables
associ\'es aux surfaces {\bf K3} '', 
``Problems in the theory of surfaces and their
classification'' (Cortona, 1988), 25--31, Sympos. Math., XXXII, Academic Press,
London, 1991.}
\lref\Beauii{A.~Beauville, ``Counting rational curves on {\bf K3} surfaces''
 alg-geom/9701019}
\lref\BeauVol{A.~Beauville, ``Vari\'et\'es Kahleriennes dont la premi\`ere
classe de Chern est nulle'', Journ. Diff. Geom. {\bf 18} (1983) n0. 4, 
755--782}
\lref\efr{B.Enriquez, B.Feigin, V. Rubtsov, "Separation of variables for
$sl_2$-Calogero-Gaudin system", Compositio Math. 110 (1998), no. 1,
1--16, q-alg/9605030}
\lref\Fr{E.~Frenkel, ``Affine algebras, Langlands duality and Bethe
ansatz. XIth International Congress of Mathematical Physics (Paris, 1994),
606--642, Internat. Press, Cambridge, MA, 1995, q-alg/9506003}
\lref\BVS{M.~Bershadsky, C.~Vafa, V.~Sadov, "D-branes and topological
theories", Nucl.Phys. {\bf B}463 (1996) 420-434}
\lref\AltKl{A.~Altman, S.~Kleiman, `` Compactifying the Picard scheme,
Adv.~in Math., 35 (1980), 50-112}
\lref\sklyanin{E. Sklyanin, ``Separation of variables. New trends''
solv-int/9505035}
\lref\vafabound{C.~Vafa, ``Instantons on D branes'', hep-th/9512078,
Nucl. Phys. {\bf B} 463 (1996) 435-442}
\lref\witbound{E.~Witten, ``Bound States Of Strings And $p$-Branes'',
hep-th/9510135,
Nucl.~Phys. {\bf B}460 (1996) 335-350}
\lref\markman{E.~Markman, ``Spectral curves and integrable systems'', 
Compositio Math. 93 (1994), no. 3, 255--290. }
\lref\yaz{S.-T.~Yau, E. Zaslow,`` BPS States, string duality and nodal curves 
on K3'', hep-th/9512121}
\lref\nikas{N.~Nekrasov, A.~Schwarz, ``Instantons on non-commutative ${\IR}^4$
and six dimensional $(2,0)$ superconformal theory'', hep-th/9802068,
Comm.Math.Phys. {\bf 198} (1998) 689-703 }
\lref\bochicchio{M.~Bochicchio, ``The large-N limit of QCD and the
collective field of the Hitchin fibration'', 
hep-th/9810015}

% Something to deal with sub-sub-sections

\def\unlockat{\catcode`\@=11}
\def\lockat{\catcode`\@=12}

\unlockat
% Something to deal with sub-sub-sections

\def\newsec#1{\global\advance\secno by1\message{(\the\secno. #1)}
\global\subsecno=0\global\subsubsecno=0\eqnres@t\noindent
{\bf\the\secno. #1}
\writetoca{{\secsym
} {#1}}\par\nobreak\medskip\nobreak}
\global\newcount\subsecno \global\subsecno=0
\def\subsec#1{\global\advance\subsecno by1\message{(\secsym\the\subsecno. #1)}
\ifnum\lastpenalty>9000\else\bigbreak\fi\global\subsubsecno=0
\noindent{\it\secsym\the\subsecno. #1}
\writetoca{\string\quad {\secsym\the\subsecno.} {#1}}
\par\nobreak\medskip\nobreak}
\global\newcount\subsubsecno \global\subsubsecno=0
\def\subsubsec#1{\global\advance\subsubsecno by1
\message{(\secsym\the\subsecno.\the\subsubsecno. #1
)}
\ifnum\lastpenalty>9000\else\bigbreak\fi
\noindent\quad{\secsym\the\subsecno.\the\subsubsecno.}{#1}
\writetoca{\string\qquad{\secsym\the\subsecno.\the\subsubsecno.}{#1}}
\par\nobreak\medskip\nobreak}

\def\subsubseclab#1{\DefWarn#1\xdef #1{\noexpand\hyperref{}{subsubsection}%
{\secsym\the\subsecno.\the\subsubsecno}%
{\secsym\the\subsecno.\the\subsubsecno}}%
\writedef{#1\leftbracket#1}\wrlabeL{#1=#1}}% Macros for boxes
\lockat

%%%%%%%%%%%%%%%%%%%%%  Rublenye bukvy   %%%%%%%%%%%%%%%%%%%%%%%%
\def\H{{\rm H}}
\def\IA{{\bf A}}
\def\IB{{\bf B}}
\def\IC{{\bf C}}

\def\IG{\relax\hbox{$\inbar\kern-.3em{\rm G}$}}
\def\IGa{\relax\hbox{${\rm I}\kern-.18em\Gamma$}}
\def\IH{\relax{\rm I\kern-.18em H}}
\def\II{\relax{\rm I\kern-.18em I}}
\def\IK{\relax{\rm I\kern-.18em K}}
\def\IL{\relax{\rm I\kern-.18em L}}
\def\IM{\relax{\rm
I\kern-.18em M\kern -.18em I}}
\def\IP{{\bf P}}
\def\IR{{\bf R}}
\def\IZ{{\bf Z}}
%%%%%%%%%%%%%%%%%%%% Calligraphic letters  %%%%%%%%%%%%%%%%%%%%%%%

\def\CD{{\cal D}}
\def\CE{{\cal E}}
\def\CF{{\cal F}}
\def\CG{{\cal G}}
\def\CH{{\cal H}}
\def\CI{{\cal I}}
\def\CJ{{\cal J}}

\def\CL{{\cal L}}
\def\CM{{\cal M}}
\def\CN{{\cal N}}
\def\CO{{\cal O}}
\def\CP{{\cal P}}

\def\CS{{\cal S}}

\def\CU{{\cal U}}
\def\CV{{\cal V}}

\def\CX{{\cal X}}

%%%%%%%%%%%%%%%%%%%%%%%%%% Derivatives  %%%%%%%%%%%%%%%%%%%%%%%%

\def\p{\partial}
\def\pb{\bar{\partial}}
%%%%%%%%%%%%%%%%%%%% letters with bar %%%%%%%%%%%%%%%%%%%%%%%%%%

\def\zb {\bar{z}}

%%%%%%%%%%%%%%%%%%%%%%%%%%% Math symbols %%%%%%%%%%%%%%%%%%%%%%%

\def\Det{{\rm Det}}

\def\Tr{\rm Tr}
\def\Id{{\rm Id}}

%%%%%%%%%%%%%%%%%%%%%%%%%%%%%%%%%%%%%%%%%%%%%%%%%%%%%%%%%%%%%%%%%
\font\manual=manfnt \def\dbend{\lower3.5pt\hbox{\manual\char127}}

\def\c{\cdot}

\def\Det{{\rm Det}}

\def\Hom{{\rm Hom}}

\def\inbar{\,\vrule height1.5ex width.4pt depth0pt}
%%%%%%%%%%
%%%%%%% Lie algebras %%%%%%%%%%%%%%%%%%%%%%

\def\Lie{{\rm Lie}}
\def\lieg{{\underline{\bf g}}}

%%%%%%%%%%%%%%%%%%%%%%%%%%%%%%%%%%%%%%%%%%%%%%%%%%%%%%%%%%%%%%%%%

% Macros for boxes

\def\boxit#1{\vbox{\hrule\hbox{\vrule\kern8pt
\vbox{\hbox{\kern8pt}\hbox{\vbox{#1}}\hbox{\kern8pt}}
\kern8pt\vrule}\hrule}}
\def\mathboxit#1{\vbox{\hrule\hbox{\vrule\kern8pt\vbox{\kern8pt
\hbox{$\displaystyle #1$}\kern8pt}\kern8pt\vrule}\hrule}}

%% ANOTHER SET OF MACROS

\def\inbar{\,\vrule height1.5ex width.4pt depth0pt}

\def\res{{\rm res}}
\def\neib{neibourghood}

\lref\hklr{Hitchin, Karlhede, Lindstrom, and Rocek,
``Hyperkahler metrics and supersymmetry,''
Commun. Math. Phys. {\bf 108}(1987)53
5}

\lref\FFR{B. Feigin , E. Frenkel, N. Reshetikhin,  Gaudin Model,
Critical Level and Bethe Ansatz, Comm. Math. Phys.  166 (1995), 27-62}
\lref\efk{P.~ Etingof, I.~ Frenkel, A.~ Kirillov, Jr.,
``Spherical functions on affine Lie groups'', Yale preprint, 1994}
\lref\etingof{P.I. Etingof and I.B. Frenkel,
``Central Extensions of Current Groups in
Two Dimensions,'' Commun. Math.
Phys. {\bf 165}(1994) 429}
%%%%%%%%%%%%%%%%%%% Kronheimer %%%%%%%%%%%%%%%%%%%%%%%%%%%%%
\lref\kronheimer{P. Kronheimer, ``The construction of ALE spaces as
hyper-kahler quotients,'' J. Dif
f. Geom. {\bf 28}1989)665}
%%%%%%%%%%%%%%%%%%% Nakajima %%%%%%%%%%%%%%%%%%%%%%%%%%%%%%%%
\lref\KN{P.~ Kronheimer and H.~ Nakajima,  ``Yang-Mills instantons
on ALE gravitational instantons,''  Math. Ann.
{\bf 288} (1990) 263}
\lref\nakajima{H. Nakajima, ``Homology of moduli
spaces of instantons on ALE Spaces. I'' J. Diff. Geom.
{\bf 40} (1990) 105; ``Instantons on ALE spaces,
quiver varieties, and Kac-Moody algebras,'' preprint,
``Gauge theory on resolutions of simple singularities
and affine Lie algebras,'' preprint.}
\lref\gncal{A.~ Gorsky, N.~ Nekrasov, ``Hamiltonian
systems of Calogero type and Two Dimensional Yang-Mills
Theory'', Nucl. Phys. {\bf B} (1994) }
\lref\gnell{A.~ Gorsky, N.~ Nekrasov, ``Elliptic Calogero-Moser System from
Two Dimensional Current Algebra'', hepth/9401021}
\lref\gnru{A.~ Gorsky, N.~ Nekrasov, ``Relativistic Calogero-Moser model as
gauged WZW theory'', Nucl.Phys. {\bf B} 436 (1995) 582-608}
\lref\dual{V.~ Fock, A.~ Gorsky, N.~ Nekrasov, V.~ Rubtsov,
``Duality in Many-Body Systems and Gauge Theories'', in preparation}
%%%%%%%%%%%
%%%%%%%%%%%%%%%% Krichever %%%%%%%%%%%%%%%%%%%%%%%%
\lref\kr{ I.~Krichever, Funk. Anal. and Appl., {\bf 12}
(1978),  1, 76-78; {\bf 14} (1980), 282-290}

\lref\beil{A.~ Beilinson, V.~ Drinfeld,
``Quantization of Hitchin's fibration and Langlands' program'',
Algebraic and geometric methods in mathematical physics (Kaciveli,
1993),
 3--7, Math. Phys. Stud., 19, Kluwer Acad. Publ., Dordrecht, 1996}
\lref\c{F.~ Calogero, J.~ Math.~ Phys. {\bf 12} (1971) 419}
\lref\ga{R. Garnier, ``Sur une classe de system\`es diff\'erentiels
Abeli\'ens
d\'eduits de la th\'eorie des \'equations lin\'eares'', Rend. del Circ.
Matematice Di Palermo, {\bf 43}, vol. 4 (1919)}
\lref\gau{M. Gaudin, Jour. Physique, {\bf 37} (1976), 1087-1098}
\lref\kks{D.~ Kazhdan, B.~ Kostant and S.~ Sternberg, Comm. on Pure and
Appl. Math., vol. {\bf XXXI}, (1978), 481-507}
\lref\op{M.~ Olshanetsky, A.~ Perelomov, Phys. Rep. {\bf 71} (1981), 313}
\lref\m{J.~ Moser, Adv.Math. {\bf 16} (1975), 197-220; }
\lref\rs{S.N.M. Ruijsenaars, H. Schneider, Ann. of Physics {\bf 170} (1986),
 370-405}
\lref\r{S.~ Ruijsenaars, Comm. Math. Phys.  {\bf 110}  (1987), 191-213 }
\lref\su{B.~ Sutherland, Phys. Rev. {\bf A5} (1972), 1372-13
76}
\lref\sch{L.~ Schlesinger, `` \"Uber eien Klasse
von Differentialsystemen
beliebiger Ordnung mit festen kritischen Punkten'',
Journal f\"ur die reine und angewandte Mathematik,
Bd. CXLI (1912), pp. 96-145}
\lref\er{B.~ Enriquez, V.~ Rubtsov, ``Hitchin systems, higher
Gaudin operators and {\sl r}-matrices'',
alg-geom/9503010,Math. Res. Lett. 3 (1996), no. 3, 343--357}

\Title{ \vbox{\baselineskip12pt\hbox{hep-th/9901089}
\hbox{HUTP-A036/98}
\hbox{ITEP-TH-36/97}}}
{\vbox{
\centerline{Hilbert Schemes, Separated Variables, and D-Branes}
}}
\bigskip
\bigskip
\centerline{A.~Gorsky$^{1}$, N.~ Nekrasov$^{2}$, V.~Rubtsov$^{3}$}

\bigskip
\centerline{\cyr $^{1,2,3}$ Institut Teoretiqesko\u\i\quad i 
E1ksperimentalp1no\u\i\quad Fiziki, 
117259, Moskva, Rossiya}
\centerline{\rm $^{2}$ Lyman Laboratory of Physics,  Harvard University, 
Cambridge MA 02138 USA}
\centerline{\rm $^{3}$ D\'epartement de Math\'ematiques, Universit\'e d'Angers,
49045, Angers, France}
\centerline{\rm $^{1,2,3}$ Institut Mittag-Leffler, Aurav\"agen 17, Djursholm, 
Sweden }
\medskip
\centerline{\tt gorsky@vitep3.itep.ru,
 nikita@string.harvard.edu, volodya@orgon.univ-angers.fr}
\bigskip\bigskip\bigskip
\lref\dualus{V.~Fock, A.~Gorsky, N.~Nekrasov, V.~Rubtsov}
\lref\bds{T.~Banks, M.~Douglas, S.~Seiberg}

We explain Sklyanin's separation of variables in geometrical terms
and construct it for Hitchin and Mukai integrable systems. 
We construct Hilbert schemes of points on $T^{*}\Sigma$ for
$\Sigma = {\IC}, {\IC}^{*}$ or elliptic curve, and on 
${\bf C}^{2}/{\Gamma}$
 and show
that their complex deformations are integrable systems of
Calogero-Sutherland-Moser type.
We present the hyperk\"ahler quotient constructions for
Hilbert schemes of points on cotangent bundles to the higher genus
curves,
utilizing the results of Hurtubise, Kronheimer and Nakajima. 
Finally we discuss the 
connections to physics of $D$-branes and string duality.

\Date{01/99}

\newsec{Introduction}

A way of  solving a 
problem
with many degrees
of freedom is to reduce it to the problem with the 
smaller
number of
degrees
of freedom.
The solvable models  allow to reduce the original system
with $N$
degrees
of freedom to $N$ systems with $1$ degree of freedom
which reduce to 
quadratures. This approach is called a separation of
variables ({\bf SoV}). Recently, 
E.~ Sklyanin came up with a ``magic recipe'' for the {\bf SoV} in the
large
class of quantum integrable models with a Lax representation
\sklyanin\sepex. The method reduces in the classical case to the
technique of separation of variables using poles of the Baker-Akhiezer 
function, which goes back to the work \oldsov, see also \krichphong\
for recent developments and more references. The
basic
strategy of this method is to look
at the Lax   eigen-vector ( which is the  Baker-Akhiezer
function) $\Psi (z, \lambda)$:
\eqn\bakakh{L(z) \Psi(z, \lambda) = \lambda (z) \Psi(z,
\lambda)}
with some choice of normalization (this is the artistic
part of the
method). The poles $z_{i}$ of $\Psi(z, \lambda)$ together with the
eigenvalues
$\lambda_{i} = \lambda(z_{i})$
are the separated variables. In all the examples studied so far
the most naive way of normalization
leads to the canonically conjugate coordinates
$\lambda_{i}, z_{i}$.

The purpose of this paper is to explain the geometry behind
the ``magic recipe'' in a broad class of examples, which
include Hitchin systems \hi, their deformations \DonEinLaz\
and many-body systems considered as their degenerations
\nhol\er.
We shall use the results of \Hurt,\nakjack,\wilson.
For a complex surface $X$ let
$X^{[h]}$ denote the Hilbert scheme of points on $X$ of
length $h$ (if $X$ is compact hyperkahler then so is $X^{[h]}$ \BeauVol).

\newsec{Hitchin  systems}

Hitchin systems can be thought of as a  generalized
many-body system. In fact,
elliptic Calogero-Moser model as well as its various
spin and some relativistic generalizations
can be
thought as of a particular degeneration of Hitchin system
\nhol,\gncal,\er. 

\subsec{The integrable system}

Recall
the general Hitchin's setup \hi. One starts with the
compact algebraic curve $\Sigma$ of genus higher then one 
and a topologically
trivial vector
bundle $V$ over
it. Let $G = SL_{N}(\IC)$, $\lieg = {\Lie}G$. The
Hitchin
system
is the integrable system on the moduli
space $\CN$ of stable Higgs bundles.  The point of
$\CN$ is the
gauge equivalence class of a pair ( an  operator $\pb_{A} = \pb +\bar A$,
a holomorphic section $\phi$ of ${\rm ad}(V) \otimes \omega^{1}_{\Sigma}$).
The holomorphic structure on $V$ is defined with the help
of $\pb_{A}$.
The symplectic structure on $\CN$ descends from the
two-form
$\int {\Tr} \delta \phi \wedge \delta \bar A$. The
integrals of
motion are the
Hitchin's hamiltonians:
$$
{\Tr} \phi^{k} \in {\H}^{0} ({\Sigma} , \omega^{k}) \approx
{\IC}^{(2k-1)(g-1)}, \quad k > 1
$$
Their total number is equal to:
$$
\sum_{k=2}^{N-1} (2k-1)(g-1) = (N^{2}-1)(g-1) = {\half}
{\rm dim}_{\IC}{\CN}
$$
Thus,
$\CN$ can be represented as fibration over
$$
{\bf B} = \bigoplus_{k=2}^{N-1} {\H}^{0} ({\Sigma} , \omega^{k})
$$
with the fibre over a generic point $b \in {\bf B}$ being
an abelian
variety $E_{b}$. This variety is identified by Hitchin
with the quotient of the Jacobian variety $J_{b}$ of the
spectral
curve $C_{b}$, defined as the divisor of zeroes of
$$
R(z, {\lambda}) = {\Det} ({\phi} (z) - {\lambda} ) \in
{\H}^{0} ({\Sigma}, \omega^{N})
$$
in $T^{*}\Sigma$. The curve $C_{b}$ has genus
$N^{2}(g-1) +1$ which
is by $g$ higher then the dimension of $E_{b}$.
In fact, $E_{b} = J_{b}/ {\rm Jac}({\Sigma})$.
The Jacobian of $\Sigma$ is embedded into the Jacobian of
$C$ as follows. The spectral curve covers $\Sigma$, hence  the
holomorphic
$1$-differentials
on $\Sigma$ are pulled back to $C_{b}$ giving the desired embedding.
An open dense subset of $\CN$ is isomorphic to $T^{*}\CM$, the cotangent
bundle to the moduli space of holomorphic stable $G$-bundles
on $\Sigma$.

The space ${\CN}$ is a non-compact integrable
system. One can
compactify
it by replacing $T^{*}{\Sigma}$ by a ${\bf K3}$ surface.
This is a
natural
deformation of the original system in the sense that the infinitesimal
{\neib} of $\Sigma$
imbedded into ${\bf K3}$ is isomorphic to $T^{*}{\Sigma}$
(since $c_{1}({\bf K3})=0$).
Instead of studying the moduli space of the gauge fields
on $\Sigma$ together with the Higgs fields $\phi$ one
studies the moduli  of torsion free sheaves,
supported on $\Sigma$ \DonEinLaz.

It turns out that this model is important in the studies
of the bound states of $D2$ branes in Type ${\II}$A string
theory compactified on $\bf K3$, which wrap a given
holomorphic curve $\Sigma$. One can think of the
bound state as of the vacuum in the gauge theory
on $\Sigma$ according to \witbound, \vafabound.
It can also be represented classically by
a smooth curve of the genus $h = N^{2}(g-1)+1$, which is
an $N$-fold cover of $\Sigma$.

In the compactified case, the curve $C$ (which is
nothing but our fellow spectral curve $C_{b}$) is
imbedded into $\bf K3$. It is also endowed with the
line bundle $\CL$ (from the point of view of Hitchin equations the
bundle is simply the eigen-bundle of $\phi$; from the
$D$-brane point of view - the single $D$-brane carries a
$U(1)$ gauge field whose  vacuum configurations are the 
flat connections) which determines a
point on its Jacobian $J_{b}$. Let us assume that ${\rm deg}{\CL} = h$.

Take the generic section of this line bundle. It has $h$
zeroes $p_{1}, \ldots, p_{h}$.
Conversely, given a set $S$ of $h$ points in $\bf K3$ there is
generically a unique curve $C$ in a given homology
class $\beta \in {\H}_{2}({\bf K3}, {\IZ})$ of genus $h$
with a line bundle $\CL$ on it, such that the curve
passes through these points and the divisor of $\CL$
coincides with $S$.

This  identifies the open dense subset of the moduli space of pairs (a
curve $C_{b}$;  a line bundle $\CL$ on it) with that in the 
symmetric power ${\rm Sym}^{h}({\bf K3})$ of the
$\bf K3$ surface itself.
The symplectic form on the moduli space is therefore
the direct sum of $h$ copies of the symplectic forms on
$\bf K3$.

Summarizing, the phase space of integrable system looks
locally as $T^{*}\CM$, where $\CM$ is the moduli
space of rank $N$ vector bundles over $\Sigma$, where
$\Sigma$ is holomorphically imbedded in
$\bf K3$. It can also be identified with the moduli space
of pairs  ( a curve $C_{b}$;  a
line bundle $\CL$ on it) where the homology class of
$C_{b}$ is $N$
times that of $\Sigma$, and the topology
$\CL$ is fixed. This identification provides
the action-angle coordinates on the phase space.
Namely, the angle coordinates are (following \hi) the linear
coordinates on the Jacobian of $C_{b}$, while
the action coordinates are the periods of ${\rm d}^{-1} \omega$
along the $A$-cycles on $C_{b}$. The last identification of the phase space
with the symmetric power of $\bf K3$ provides the
$\bf SoV$ in the sense of Sklyanin.
Notice the similarity of our description to his ``magic recipe''.

\subsec{Separation of variables in Hitchin system}

Let us present an explicit realization of the separation
of variables in the Hitchin system.
Let $\Sigma$ be a compact smooth genus $g$ algebraic curve
and $p$ be the projection $p: T^{*}\Sigma \to \Sigma$.
Let $V$ be complex vector bundle over $\Sigma$ of rank $N$ and
degree $k$. Consider the moduli space $\CM_{N,k}$ of the
semi-stable holomorphic structures $E$ on $V$. It can be identified
with the quotient of the open subset of the space of
$\pb_{A}$
operators acting on the sections of $V$ by
the action of the gauge group $\pb_{A} \to g^{-1}\pb_{A} g$.
The complex dimension of $\CM_{N,k}$ is given by the
Riemann-Roch formula
\eqn\dmn{{\rm dim}_{\IC} {\CM}_{N,k} = N^{2}(g-1) + 1 := h}
Explicitly, the points in $T^{*}\CM_{N,k}$ are the equivalence
classes of pairs $(E, \Phi)$ where $E$ is the holomorphic
bundle on $\Sigma$ and $\Phi$ is the section of
${\rm End}(E) \otimes \omega_{\Sigma}$.
The map $\pi$ is given by the formula:
\eqn\htmp{\pi (E, \Phi) = \{ {\Tr} \Phi,
{\Tr} \Phi^{2}, \ldots, {\Tr} \Phi^{N} \} }
Let $\CH$ denote the $h$-dimensional vector space $\IC^{h}$:
$$
\CH  = \oplus_{l=1}^{N} {\H}^{0} ( \omega_{\Sigma}^{\otimes l} )
$$
N.~Hitchin shows that the partial compactification of the
cotangent bundle $T^{*}\CM_{N,k}$
is the algebraicaly  integrable system, i.e.
there exists a holomorphic map
$$
\pi: T^{*}\CM_{N,k} \to \CH
$$
whose fibers are abelian varieties which are Lagrangian with respect to the
canonical symplectic structure on $T^{*}
\CM_{N,k}$. The generic fiber is compact.
Geometric separation of variables in Hitchin system
is the content of the following:
\bigskip
\noindent
{\bf Theorem.} There exists a birational map 
$$
\varphi : T^{*}\CM_{N,k} \to (T^{*}\Sigma)^{[h]},
$$ 
which is a symplectomorphism of the open dense subsets.

\noindent
{\bf Remark.} The open dense set in $X^{[h]}$ coincides with
$(X^{h} - \Delta) /{\CS}_{h}$ where $\Delta$ denotes the union
of all diagonals and $\CS_{h}$ is the symmetric group.
The theorem implies that on this dense set one can introduce
the coordinates $\{ (z_{i}, \lambda_{i}) \}$, where $z_{i} \in \Sigma$,
$\lambda_{i}  \in T^{*}_{z_{i}} \Sigma$, such that the
symplectic form in these coordinates have the separated form:
\eqn\sepr{\Omega = \sum_{i=1}^{h} \delta \lambda_{i} \wedge \delta z_{i}}
The coordinates $(\lambda_{i}, z_{i})$ are defined up to
permutations.

\bigskip
\noindent
{\bf Proof.}
First we construct $\varphi$ and prove that this map is
biholomorphic. In order to do that we need to choose a specific
$k$. As the moduli spaces $\CM_{N,k}$ are isomorphic
for different $k$ (although not canonically!) this is not
a problem. Fix the pair $(E,\Phi)$.
Consider the spectral curve $C \subset T^{*}\Sigma$
defined as the zero set of the characteristic polynomial
\eqn\chr{P ({\lambda}, z)  = {\Det} (\Phi(z) - \lambda) \in {\Gamma}
(\omega_{\Sigma}^{\otimes N}) }
Its genus equals $h$ as can be seen from the adjunction formula
or from Riemann-Hurwitz formula.   The pullback $p^{*}E$ restricted to
$C$
contains
the line sub-bundle $L$ of the eigen-lines of $\Phi$. Our choice of
$k$ is such that the degree of $L$ equals $h$. Generically
$L$ has unique up to a multiple non-vanishing section $s$.
Its zeroes $l_{1}, \ldots , l_{h}$
determine $h$ points on $C$ and therefore on $T^{*}\Sigma$.
Clearly they are determined uniquely up to
permutations. This allows us to define:
\eqn\dfnt{{\varphi} (E, \Phi) = ( l_{1}, \ldots, l_{h})}
Now let us prove that the image of $\varphi$ coincides with
$(T^{*}\Sigma)^{[h]}$. We do it at the level of the open
dense sets.
Fix the set
of points
$(l_{1} = \left(\lambda_{1}, z_{1}), \ldots, l_{h}
= (\lambda_{h}, z_{h}) \right)$
in $T^{*}\Sigma$.
Let $\omega_{\alpha}^{(j)}$ denote a basis in the space of
holomorphic $j$-differentials on $\Sigma$. For given $j$
the index $\alpha$ runs from $1$ to $(2j-1)(g-1)$ for $j \geq 2$,
to $g$ for $j=1$ and to $1$ for $j=0$. Consider the
space $\CV$ of sections of $p^{*}\omega_{\Sigma}^{\otimes N}$.
The vectors of $\CV$ are:
\eqn\polynm{P(\lambda, z) = \sum_{j=0}^{N} \sum_{\alpha} V_{j, \alpha}
\omega_{\alpha}^{(j)} (z) \lambda^{N-j}}
The dimension of $\CV$ equals $h+1$. Consider the subspace of
$P$'s such that $P(\lambda_{i}, z_{i}) = 0$ for any $i=1, \ldots, h$.
It is one-dimensional if the points $l_{1}, \ldots, l_{h}$ are
distinct. In that case choose any $P$ in this one dimensional subspace
and consider the curve $C$ defined by the equation $P(\lambda,z) = 0$
through the points $l_{1}, \ldots, l_{h}$. They form a divisor of the
unique line bundle $L$ on $C$ of degree $h$ which can be considered as
a point of $(T^{*}\Sigma)^{[h]}$.
Then we can consider a direct image $p_{*}L$ which is a vector bundle
over $\Sigma$ and a Higgs form $\Phi : p_{*}L \to
p_{*}L\otimes\omega_{\Sigma}$ which is nothing but the multiplication by the
elements $\lambda \in C$. Hence we obtain desired
one-to-one correspondence  at the level of the open dense subsets.

In fact, one can go further and extend this isomorphism to the loci of
positive codimension. We don't need this for the birational isomorphism
here, yet let us sketch a direction of thought. A point $\ell$ in 
$(T^{*}\Sigma)^{[h]}$ determines a complex line in the space $\CV$. The divisor
of the non-zero element $P$ of this line defines a ``curve'' $C$. Now, the
point $\ell$ could be identified with the torsion free rank one sheaf on 
$T^{*}{\Sigma}$. For those $C$ which are smooth the restriction of the
sheaf becomes a line bundle ${\CL}$.

Now let us prove the equality of symplectic forms on $T^{*}\CM_{N,k}$
and on $(T^{*}\Sigma)^{[h]}$. We show
it at the generic
point of $(T^{*}\Sigma)^{[h]}$ corresponding to the set
of distinct points $(l_{1}, \ldots, l_{h})$. To this end recall
the construction of the symplectic form on $T^{*}\CM_{N,k}$
in terms of the data $(C,L)$:
Let $A_{a}$ be a choice of the basis of $A$-cycles on $C$. The
Jacobian ${\rm Jac}(C) ={\H}^{(0,1)}(C, \IC) / {\H}^{1}(C,\IZ) $
has the local linear coordinates $\varphi_{b}$
associated to $A_{a}$.  The differentials $d\varphi_{b}$ are
naturally identified with the holomorphic one-differentials
on $C$. They are normalized in
such a way that:
\eqn\nrmls{\int_{A_{a}} d\varphi_{b} = \delta^{a}_{b}.}
Introduce the coordinates $I^{a}$:
\eqn\actncr{I^{a} = \int_{A_{a}} \lambda dz.}
Then:
\eqn\actn{\omega = \sum_{a=1}^{h} \delta I^{a} \wedge \delta \varphi_{a}}

To make contact with  \sepr\ we recall the Abel map:
Let $\Omega_{a}$ be the basis in the space of holomorphic $1$-differentials
on $C$ which obeys:
\eqn\nrmlzii{\int_{A_{a}} \Omega_{b} = \delta^{a}_{b}.}
Then
$$
\varphi_{a} = \sum_{i=1}^{h} \int_{l_{*}}^{l_{i}} \Omega_{a}
$$
Notice that the normal bundle $N_{T^{*}\Sigma \vert C}$ to $C$ is
isomorphic to $T^{*}C$. 
The deformed curve $\tilde C$ can be identified with the
holomorphic section $p(x)dx$ of $T^{*}C$.
It can be expanded as follows:
$$
p(x)dx = \sum_{a=1}^{h} p^{a} \Omega_{a}, \quad p^{a} \in \IC
$$
Using \nrmlzii\ we get:
\eqn\nrmlziii{p^{a} = \oint_{A_{a}} p(x)dx = I^{a}}
Let $(p_{i}, x_{i})$, $i=1, \ldots, h$ be a set of distinct
points in $T^{*}C$. Let $C_{a, i} =  \Omega_{a} (x_{i}) \in T^{*}_{x_{i}}C$.

\noindent
{\bf Lemma.}
\eqn\cdx{\delta \varphi_{a} = \sum_{i=1}^{
h} C_{a,i} \delta x_{i}}

\noindent
{\bf Proof of the Lemma.}
$$
\delta\varphi_{a} =
\sum_{i=1}^{h} \delta \int_{l_{*}}^{(p_{i}, x_{i})} \Omega_{a} =
\sum_{i=1}^{h} \Omega_{a}(x_{i}) \delta x_{i}
$$
Thus,
$$
\sum_{i=1}^{h} \delta p_{i} \wedge \delta x_{i} =
\sum_{a=1}^{h} \sum_{i=1}^{h} \delta p^{a} C_{a,i} \delta x_{i} =
\sum_{a=1}^{h} \delta I^{a} \wedge \delta \varphi_{a}
$$

The theorem is proven.
\bigskip
\noindent
{\bf Remark.} J.~Hurtubise in \Hurt\ gave the
 pure algebro-geometric proof
of the main part of the theorem. The basic
motivation of our remarks is
that some of our arguments looks
more direct and more in the
spirit of the approach of
integrable systems.

\newsec{Gaudin model}

Consider the space
\eqn\phs{
\left( \CO_{1} \times \ldots \times \CO_{k} \right) // G}
where $\CO_{l}$ are the complex
coadjoint orbits of $G = {\rm SL}_{N}({\IC})$ 
and the symplectic quotient is taken with respect to
the diagonal action of $G$.

This moduli space parameterizes Higgs pairs 
on ${\IP}^{1}$ with singularities at the marked points
$z_{i} \in {\IP}^1, \quad i =1, \ldots, k$. This is a natural
analogue of the Hitchin space for genus zero.
Concretely, 
the connection to the bundles on $\IP^{1}$ comes about as follows:
consider the moduli space of Higgs pairs: $(\pb_{A}, \phi)$
where $\phi$ is a {\it meromorphic} section
of ${\rm ad} (V) \otimes \CO(-2)$, with the restriction that
${\res}_{z = z_{i}} \phi \in \CO_{i}$.
The moduli space is isomorphic to \phs.
This space is integrable system,
studied in \ga\gau. Indeed, consider the solution
to the equation
\eqn\mmp{\pb_{A}\phi = \sum_{i} \mu_{i}^{c} \delta^{(2)}(z - z_{i})}
in the gauge where $\bar A=0$ (such gauge exists for stable bundles
on $\IP^{1}$ due to Grothendieck's theorem).
We get:
\eqn\lopr{\phi(z) = \sum_{i} {{\mu_{i}^{c}}\over{z - z_{i}}}}
provided that $\sum_{i} \mu_{i}^{c} =0$
and is defined up to a global conjugation by an element
of $G$ hence the Hamiltonian reduction in \phs.
Now, consider the following polynomial:
\eqn\spctr{{\Det}\left(\lambda - \phi(z)\right) = \sum_{i,l} A_{i,l}
\lambda^{i} z^{-l}}
It is an easy count to check that the number of functionally
independent coefficients $A_{i,l}$ is precisely
equal to
$$
k \left( {{N (N-1)}\over{2}} \right) + 1 - N^{2}
$$
Now let us treat explicitly the case $N=2$. In this case the coadjoint
orbits
$\CO_{i}$ can be explicitly described as the surfaces in $\IC^{3}$ given
by the equations:
\eqn\orbts{\CO_{i} : Z_{i}^{2}+ X_{i}^{+}X_{i}^{-} = \zeta_{i}^{2}}
with the symplectic forms:
\eqn\smpl{\omega_{i} = {{dZ_{i} \wedge dX_{i}^{+}}\over{X_{i}^{+}}}}
and the complex moment maps:
\eqn\mmcmp{\mu_{i}^{c} = \pmatrix{Z_{i} &   X_{i}^{+}  \cr
X_{i}^{-}  & - Z_{i}\cr} }
The phase space of our interest is $\CP = \times_{i=1}^{k} \CO_{i} //
SL_{2}$.
It is convenient to work with a somewhat larger space $\CP_{0} =
\times_{i=1}^{k}
\CO_{i} / \IC^{*}$, where $\IC^{*} \in SL_{2}(\IC)$ acts as follows:
$$
t : \left( Z_{i}, X_{i}^{+}, X_{i}^{-} \right) \mapsto
\left( Z_{i}, t X_{i}^{+} , t^{-1} X_{i}^{-} \right)
$$
The moment map of the torus $\IC^{*}$ action is simply $\sum_{i} Z_{i}$.
The complex dimension of $\CP_{0}$ is equal to $2(k-1)$.
The Hamiltonians are obtained by expanding the
quadratic invariant:
\eqn\str{\eqalign{& T(z) = {\half} {\Tr} \phi(z)^{2}, \qquad
\phi(z)  = \sum_{i}
{{\mu_{i}^{c}}\over{z-z_{i}}} \cr
& T(z) = \sum_{i}  {{\zeta_{i}^{2}}\over{(z-z_{i})^{2}}} +
\sum_{i} {{H_{i}}\over{z-z_{i}}} \cr
& H_{i} = {\half} \sum_{j \neq i} {{X_{i}^{+}X_{j}^{-} + X_{i}^{-}X_{j}^{+}
+ 2 Z_{i} Z_{j} }\over{z_{i} -
z_{j}}}\cr}}
The separation of variables proceeds in this case as follows:
write $\phi(z)$ as
$$
\phi(z) = \pmatrix{ h(z) & f(z) \cr e(z) & - h(z) \cr}.
$$
Then Baker-Akhiezer function is given explicitly by:
\eqn\baf{\Psi(z) = \pmatrix{\psi_{+} \cr \psi_{-} \cr},
\quad \psi_{+} = f, \quad \psi_{-} = \sqrt{h^{2} + ef} - h }
and its zeroes are the roots of the equation
\eqn\rts{f(p_{l}) = 0 \Leftrightarrow  \sum_{i}
{{ X_{i}^{+}}\over{p_{l} - z_{i}}} = 0, \quad l = 1, \ldots, k-1}
The eigen-value $\lambda(p)$
of the Lax operator $\phi$ at the point $p$  is most easily
computed using the fact that $T(z) = \lambda(z)^{2}$.
Hence, $\lambda_{l} = \sum_{i} {{Z_{i}}\over{p_{l} - z_{i}}}$,
\eqn\trsnf{\eqalign{
& X_{i}^{+} = u {{P(z_{i})}\over{Q^{\prime}(z_{i})}},  \quad u \in \IC\cr
& Z_{i} = \sum_{l} {{\lambda_{l}}\over{z_{i} - p_{l}}}
{{Q(p_{l})P(z_{i})}\over{Q^{\prime}(z_{i})P^{\prime}(p_{l})}} \cr
& \qquad P(z) = \prod_{l=1}^{k-1} ( z - p_{l})\cr
& \qquad Q(z) = \prod_{i=1}^{k} (z - z_{i}) \cr}}
The value of $u$ can be set to $1$ by the $\IC^{*}$ transformation.
The $({\lambda_{l}}, p_{l})$'s are
the gauge invariant coordinates on $\CP_{0}$. They are
defined up to a permutation.
It is easy to check that the restriction of the symplectic form
$\sum \omega_{i}$ onto the set $\sum_{i} Z_{i} = 0$
is the pullback of the form
\eqn\seprfrm{\sum_{l=1}^{k-1} d \lambda_{l} \wedge d p_{l}.}
In the last section we present the quantum analogue of this separation
of variables.

\newsec{Many-Body Systems: Rational and Trigonometric Cases}

In this section we study the Hilbert scheme of points on
$S = {\IC}^2$, $S = {\IC}^{2} /{\Gamma}$ for ${\Gamma} \approx {\IZ}_{N}, 
{\IZ}$. We show that $S^{[v]}$ has a complex deformation $S^{[v]}_{\zeta}$
and that each $S^{[v]}_{\zeta}$ is an 
integrable model including the complexification
of Sutherland model \su. 

\subsec{Points on ${\IC} \times {\IC}$}

Let us start with ${\IC}^2$. As  is well-known \naklect\ the Hilbert scheme
of points on ${\IC}^{2}$ has ADHM-like description: it is the set of 
stable triples $(B_1, B_2, I)$, $I \in V \approx {\IC}^{v}, B_{1}, B_{2} \in 
{\rm End}(V)$, $[B_{1}, B_{2} ] = 0$ modulo the action of ${\rm GL}(V)$:
$(B_{1}, B_{2}, I) \sim (g B_{1} g^{-1}, g B_{2} g^{-1}, g I)$ for 
$g \in {\rm GL}(V)$. 
Stability means that by acting on the vector $I$ by arbitrary polynomials
in $B_{1},B_{2}$ one can generate the whole of $V$.

The meaning of the vector $I$ and the operators $B_{1}, B_{2}$ is the following.
Let $z_{1}, z_{2}$ be the coordinates on ${\IC}^2$. Let $Z$ be
a zero-dimensional subscheme of ${\IC}^2$ of length $v$. It means that 
the space ${\H}^{0}({\CO}_{Z})$ 
of functions on $Z$ which are the restrictions of holomorphic
functions on ${\IC}^2$ has dimension $v$. Let $V$ be this space of
functions. Then it has the canonical vector $I$ which is the constant
function $f = 1$ restricted to $Z$ and the natural action of
two commuting operators: multiplication by $z_1$ and by
$z_2$, which are represented by the operators $B_1$ and $B_2$. 
\noindent
Conversely, given a stable 
triple $(B_{1}, B_{2}, I)$ the scheme $Z$, or, rather
the corresponding ideal ${\CI}_{Z} \subset {\IC} [ z_1, z_2 ]$
is reconstructed as follows: $f \in {\CI}_{Z}$ iff $f (B_{1}, B_{2} ) I = 0$.

Now let us discuss the notion of stability. In the Geometric Invariants
Theory  (GIT) the notion of a stable triple $(B_{1}, B_{2}, I)$ 
would be the following:
{\it there exists a holomorphic function $\psi$
on the space of all triples 
$({\tilde B}_{1},{\tilde B}_{2}, {\tilde I})$ such that:}

\item{1.} {\it For any   $g \in {\rm GL}(V)$, \quad
$ \psi ( g {\tilde B}_{1} g^{-1}, g {\tilde B}_{2} g^{-1}, 
g {\tilde I}) = {\rm det}(g) \psi 
({\tilde B}_{1}, {\tilde B}_{2},  {\tilde I})$ }

\item{2.} $\psi (B_{1}, B_{2}, I) \neq 0$.

Let us show the equivalence of the two definitions of stability.
Choose any $v$-tuple $\vec f$ 
of polynomials $f_{1}, \ldots, f_{v} \in {\IC}
[ z_{1}, z_{2}]$. 
Choose any non-zero element 
$\omega \in \left( \Lambda^{v} {\IC}^{v} \right)^{*}$.  Define a function 
\eqn\ttau{\tau_{\vec f} ({\tilde B}_{1}, {\tilde B}_{2} , {\tilde I}) = 
\omega \left( f_{1} ({\tilde B}_{1},{\tilde B}_{2}) {\tilde I} \wedge
 \ldots \wedge 
f_{v}({\tilde B}_{1} , {\tilde B}_{2}) {\tilde I} \right)}
Clearly it obeys the property $\bf 1.$ If the triple $(B_{1}, 
B_{2}, I)$ is stable
in the sense of the first definition then there exist a $v$-tuple
$\vec f$ for which the vectors $ f_{1} (B_{1}, B_{2}) I, 
\ldots, f_{v}(B_{1} , B_{2}) I $ form a basis in ${\IC}^{v}$
and therefore $\tau_{\vec f} (B_{1}, B_{2}, I) \neq 0$. Conversely, 
if  $\tau_{\vec f} (B_{1}, B_{2}, I) = 0$ for any $\vec f$ then
the span $S$ of $\lbrace {\IC} [ B_{1}, B_{2} ] I \rbrace$ is strictly less
then $V$. On the other hand $S$ is an invariant subspace.

Now let us discuss another aspect of the space $\left( {\IC}^{2} \right)^{[v]}$.
It is symplectic manifold. To see this let us start with the space
of quadruples, $(B_{1}, B_{2}, I, J)$ with $B_{1}, B_{2}, I$ as above
and $J \in V^{*}$. It is a symplectic manifold with the symplectic form
\eqn\sone{\Omega = {\Tr}\left[  {\delta} B_{1} \wedge {\delta} B_{2} + 
\delta I \wedge {\delta} J  \right]}
which is invariant under the naive action of $G = {\rm GL}(V)$. The moment map
for this action is 
\eqn\mone{\mu = [B_{1}, B_{2} ] + IJ \in {\rm Lie}G . }
Let us perform the Hamiltonian reduction, that is:

\noindent
take the zero level set of $\mu$, choose a subset of stable points in the 
sense of GIT and take the quotient
of this subset with respect to $G$. One can show \naklect\
that the stability implies that $J =0$ and therefore the moment equation reduces
to the familiar $[B_1, B_2]=0$.

Moreover, $\left( {\IC}^{2} \right)^{[v]}$ is an integrable system. Indeed,
the functions ${\Tr} B_{1}^{l}$ Poisson-commute and are
functionally independent for $l= 1, \ldots, v$.

It turns out that $\left( {\IC}^{2} \right)^{[v]}$ has an interesting
complex deformation which preserves its symplecticity and 
integrability. Namely, instead of $\mu^{-1}(0)$ in the reduction one should take
${\mu}^{-1} ( \zeta \cdot {\Id} )$ for some $\zeta \in {\IC}$.
Now $J \neq 0$. The resulting quotient $S_{\zeta}^{[v]}$ no longer parametrizes
subschemes in ${\IC}^2$ but rather sheaves on a non-commutative ${\IC}^{2}$,
that is the ``space'' where functions are polynomials in $z_1, z_2$ with the
commutation relation $[z_1, z_2] = \zeta$ (see \nikas\ for more details).
Nevertheless, the quotient itself is a perfectly well-defined symplectic
manifold with an integrable system on it: 
the functions $H_{l}= {\Tr} B_{1}^{l}$ 
still Poisson-commute and are
functionally independent for $l= 1, \ldots, v$. On the dense open subset
of $S^{[v]}_{\zeta}$ where $B_{2}$ can be diagonalized:
$B_{2} = {\rm diag} \left( q_1, \ldots, q_{v} \right)$ the Hamiltonians
$H_{1}, H_{2}$ can be written as follows:
\eqn\hmlt{H_{1} = \sum_{i} p_{i}, \quad H_{2} = \sum_{i} p_{i}^{2} +
\sum_{i < j} {{\zeta^2}\over{\left( q_{i} - q_{j} \right)^2}}}
where $p_{i} = \left( B_{1} \right)_{ii}$. These Hamiltonians
describe a collection of indistinguishable particles on a (complex)
line with a pair-wise potential interaction $1\over{x^2}$. This system
is called rational Calogero model \c. It is shown in \wilson\ that
the space $S^{[v]}_{\zeta}$ can be used for compactifying the Calogero
flows in the complex case and moreover that the same compactification
is natural in the KdV/KP realization of Calogero flows \amm\kr.
 
\subsec{Rounding off to ${\IC} \times {\IC}^{*}$ and to 
${\IC}^{*} \times {\IC}^{*}$}

Now let $z_{1}, z_{2}$ be the coordinates on ${\IC} \times {\IC}^{*}$, i.e.
$z_{2} \neq 0$. Then the description of the previous section is still
valid except that $B_{2}$ must be invertible now. So in this case the Hilbert
scheme of points is obtained by a complex Hamiltonian reduction
from the space $T^{*}\left( G \times  V \right)$. We skip all the details
as they are well-known by now (the complex analogue of the real reduction
studied in \kks\op\ can be found for example in \nhol\ ). 
The moment map in our
notations will be:
\eqn\mtwo{\mu = B_{2}^{-1} B_{1} B_{2} - B_{1} + IJ }
which corresponds to the symplectic form:
\eqn\stwo{\Omega = \delta {\Tr} \left[ 
B_{1} B_{2}^{-1} \delta B_{2} + I \delta J \right]}
The reduction at the non-zero level $\mu = \zeta \cdot {\Id}$ leads to the 
the complex analogue of either 
Sutherland \su\  or rational Ruijsenaars model \r\rs. In the former case
$H_{2} = {\Tr} B_{1}^{2}$ while in the latter
$H_{rel} = {\Tr} \left( B_{2}  + B_{2}^{-1} \right)$. On the open dense
subset where  $B_{2}$ diagonalizable:
$B_{2} = {\rm diag} \left( \exp ( 2\pi i q_{1}) , \ldots, \exp ( 2\pi i q_{v} )
\right)$ the Hamiltonian $H_{2}$ equals:
\eqn\htwo{H_{2} = \sum_{i} p_{i}^2 + 
\sum_{i < j} {{\zeta^2}\over{{\rm sin}^{2}
\left( \pi\left( {q_{i} - q_{j}} \right)\right)}}}
The expression for $H_{rel}$ can be found in \nhol.

Finally, if both $B_{1}$ and $B_{2}$ are invertible then
we get the Hilbert scheme of points on ${\IC}^{*} \times {\IC}^{*}$.
Its complex deformation is a bit more tricky, though. It turns out that
it can be obtained via {\it Poisson reduction} of 
$G \times G \times V \times V$. The integrable system
one gets in this case is the trigonometric case of
Ruijsenaars model \gnru.

\subsec{ALE models}

Slightly generalizing the results of
\kronheimer\KN\nakajima\ one may easily present
the finite-dimensional symplectic quotient construction of
the Hilbert scheme of points on $T^{*}{\IP}^{1}$.
Here it is:
Take $\CV = \IC^{v}$, $A = T^{*}{\Hom}(\CV , \CV)$,
$\tilde A = A \oplus A$ and $X = \tilde A \oplus T^{*}({\Hom}(\CV, \IC))$.
The space $X$ is acted on by the group $\CG = GL(V) \times GL(V)$.
The maximal compact subgroup $\CU$ of $\CG$ preserves the hyperkahler
structure of $X$.
The hyperkahler quotient of $X$ with respect to ${\CU}$ is
the Hilbert scheme of points
on $T^{*}\IP^{1}$ of length $v$.

This space is an integrable system. We shall prove it in a slightly more
general setting. Namely, let $S$ be
the deformation of the orbifold $\IC^{2} / {\IZ}_{k}$, where
the generator $\omega = e^{{2\pi i}\over{k}}$ of $\IZ_{k}$ acts as
follows: $(z_{0}, z_{1}) \mapsto (\omega z_{0}, \omega^{-1} z_{1})$.

\noindent
{\bf Theorem.} $S^{[v]}$ is a holomorphic integrable system.

\noindent
{\bf Proof.}

The space $S^{[v]}$ can be described as a hyperkahler quotient.
Let us take $k+1$ copy of the space $\IC^{v}$, and denote the $i$'th
vector space as $V_{i}$, $i = 0, \ldots, k$. Let us consider the
space
$$
X = \bigoplus_{i=0}^{k} {\rm Hom}(V_{i}, V_{i+1}) \oplus
{\rm Hom}(V_{i+1}, V_{i}) \bigoplus {\rm Hom}( W, V_{0}) \oplus
{\rm Hom} (V_{0}, W),
$$
where $k+1 \equiv 0$.
The space $X$ has a natural hyperkahler structure, in particular
it has a holomorphic symplectic form:
\eqn\hlsmpl{\omega = {\Tr} \delta I \wedge \delta J + \sum_{i=0}^{k}
{\Tr} \delta B_{i,i+1} \wedge
\delta B_{i+1, i} }
where $B_{i,j} \in {\rm Hom}(V_{j}, V_{i})$, $I \in {\rm Hom} (W, V_{0})$,
$J \in {\rm Hom} (V_{0}, W)$.
The space $X$ has a natural symmetry group $G = \prod_{i=0}^{k} U(V_{i})$
which  acts on $X$ as:
$$
B_{i,j} \mapsto g_{i} B_{i,j} g_{j}^{-1}, \quad I \mapsto g_{0} I,
\quad J \mapsto J g_{0}^{-1}
$$
The action of the group $G$ preserves the hyperkahler structure of $X$.
The complex moment map has the form:
\eqn\cmplxmm{\mu_{i} = B_{i,i+1} B_{i+1,i} - B_{i,i-1} B_{i-1,i} +
\delta_{i,0}
IJ}
The space $S^{[v]}$ is defined as a (projective) quotient of
$\mu^{-1}(0)$ by the
action of the complexified group $G$, which we denote
as $G_{c}$.
There is a deformation $S^{[v]}_{\zeta}$ which depends
on $k$ complex
parameters $\zeta_{0}, \ldots, \zeta_{k}, \sum_{i} \zeta_{i}=0$
defined as
\eqn\dfrms{S^{[v]}_{\zeta} = \cap_{i} \mu_{i}^{-1}(\zeta_{i} \Id) / G_{c}}
Now we present the complete set of Poisson-commuting functions on
$S^{[v]}$:
define the ``monodromy'':
\eqn\mndr{\IB_{0} = B_{0,1} B_{1,2} \ldots B_{k, 0}}
which transforms under the action of $G$ in the adjoint representation.
The invariants
\eqn\invrts{f_{l} = {\Tr}  \IB^{l}_{0}, \quad l = 1, \ldots, v}
clearly Poisson-commute on $X$, are gauge invariant and therefore descend
to the commuting functions on $S^{[v]}$ (and to $S^{[v]}_{\zeta}$ as well).
The functional independence is easily checked on the dense open
set where $S^{[v]}$ can be identified with the symmetric product of $S$'s.

\newsec{Elliptic models}

\subsec{Hilbert scheme of points on $T^{*}E$}

Let $E$ be the ellipic curve, $E^{\vee} \approx {\rm Jac}(E) = 
{\rm Pic}_{0}(E)$. For $t \in E^{\vee}$ let $L_{t}$ denote the corresponding 
line bundle. In particular, let  $0 \in E^{\vee}$
be the trivial line bundle: $L_{0} = \CO$. Let $S = T^* E$, 
let $(z_{1}, z_{2})$ be the coordinates 
on the universal cover of $S$, $z_{2} \in T^{*}$,
$\pi: S \to E$
be the projection. 

Let $Z \in S^{[v]}$. For each $t \in E^{\vee}$ let
$V_{t}$ be the space ${\H}^{0} \left( Z , {\pi}^{*}L_{t}\vert_{Z} \right)$.
The spaces $V_{t}$ form a holomorphic rank $v$ bundle $\CE$ over $E^{\vee}$. Let
$\phi: V \to V \otimes \Omega^1_{E^{\vee}}$ be the operator
which multiplies a section of ${\pi}^{*}L_{t}$ by a covector (thus 
we get a map from $V \times \Omega^1_{E}$ to $V$ which  is
the same as $ V \to V \otimes \Omega^1_{E^{\vee}}$). The fiber at $t=0$
contains a distinguished vector $I$ which is the image of 
$1 \in {\H}^{0}(Z, {\pi}^{*} {\CO} \vert_{Z})$. The triple $({\CE}, \phi, I)$
is stable in the following sense:
\noindent
{\it There exists no holomorphic subbundle ${\CF}$ of ${\CE}$, such that
$\phi ( {\CF} ) \subset {\CF} \otimes \Omega^1_{E^{\vee}}$ and
$I \in {\CF}_{0}$}.

\noindent
This is an appropriate counterpart of the notion of  a stable Higgs pair
\hid\ in the case of genus one.

Let $Z \in S^{[v]}$. Consider the cover $p: {\IC} \times {\IC} \to T^* E$. Let 
us lift $Z$ to 
the covering space. The space $V$  of functions on 
$p^{*}Z$ is ${\IZ} \oplus {\IZ}$-module. By Fourier transform we may view
this space as a space of functions on a two-torus with values
in a $v$-dimensional vector space. Let $t, \bar t$ be the coordinates
on the torus. The elements of the $v$-dimensional space $V_{t,\bar t}$
are the functions on $p^{*}Z$ which transform as follows:
\eqn\trsnf{f (z_1 + m + n \tau, z_2) = \exp {{2\pi i}\over{\tau - \bar\tau}}
 \left( m \left( t - \bar t\right) 
+ n \left( \tau \bar t - \bar \tau t \right) \right) f(z_1, z_2).}
Clearly, the function $f \equiv 1$ belongs to $V_{0,0}$. It is represented
by the vector $I \in V_{0,0}$. 

To reconstruct a scheme $Z$ given a stable triple $( \phi, \bar\partial
+ \bar A, I)$
one can go to a gauge where both $\phi$ and $\bar A$ are constant
($t$-independent). Then the matrices $(\phi, \bar A)$ commute and
together with $I$ form a stable triple
suitable for defining a subscheme of ${\IC}^2$ of length $v$. In fact, the
gauge where $\phi$ and $\bar A$ are constant  allows
extra gauge transformations which make the support of the subscheme
of ${\IC}^2$ invariant under the action of
${\IZ} \oplus {\IZ}$ by translations. The simplest yet instructive
case is that of the diagonalizable $\phi, \bar A$. 

\noindent
One can do more. The Hilbert scheme of points on $T^{*}E$
can be endowed with a hyperkahler structure.
To see this  we start with the space $\CX$
of the pairs $(A, \Phi)$ - where $A$ is $U(N)$ gauge field on the torus $E$
and $\Phi$ is the adjoint-valued one-form.
The space $\CX$ is hyperkahler and the hyperkahler
structure is preserved by the action of the gauge group $\CG$.
The latter may also act via evaluation at some points $t_{1}, \ldots, t_{s}$
on some finite-dimensional hyperkahler manifolds $\CO_{1}, \ldots, \CO_{s}$.
Let $\vec\mu_{k}$ be the hyperkahler moment map for $\CO_{k}$.
Let us consider the hyperkahler reduction of the space
$\CX \times \CO_{1} \times \ldots \CO_{s}$  with respect to $\CG$.
We first impose the hyperkahler moment map equations:
\eqn\hprklm{\eqalign{\pb \Phi_{t} + [ A_{\bar t} , \Phi_{t} ] =
\sum_{k=1}^{s} \mu_{k}^{c} \delta^{(2)} ( t - t_{k}) \cr
F_{t\bar t} + [\Phi_{t} , \Phi_{\bar t} ] =
\sum_{k=1}^{s} \mu_{k}^{r} \delta^{(2)} ( t - t_{k}) \cr}}
These equations are the genus one 
analogues of Hitchin's self-duality equations \hid.
They were studied from the point of view of complex reduction in
\nhol\gnell\markman. They also appeared in the study of intersecting
D-branes on tori \setkap (see also \taylor\ for the general discussion
of D-branes on tori), in the attempts
to find appropriate field parameterization of the Yang-Mills
theory \bochicchio. The case of our interest here is $s=1$, $\CO_{1} = 
T^{*}\IC\IP^{N-1}$.
The complex reduction in this case has been studied in \gnell\ where it
was shown that the solutions to the equations \hprklm\ (in fact of their
deformation) form a holomorphic integrable system, which turns
out to be elliptic Calogero-Moser system. The latter describes
the system of non-relativistic
particles on elliptic curve $E$, which pairwise interact via potential
$\zeta_{c}^2 \wp( z_{i} - z_{j})$ where $\zeta$ is a the period of a holomorphic
symplectic form on $\CO_{1}$ (see below).

The space $\CO_{1}$ in turn can be described as a hyperkahler quotient
of $\IC^{2N}$ with respect to the action of $U(1)$ which has charges
$(+1,-1)$ on $\IC^{N} \oplus \IC^{N}$.
Let us denote the elements of $\IC^{N} \oplus \IC^{N}$ as $I \oplus J$,
where $J$ and $I$ are the row and the column vectors respectively.
Then the equations \hprklm\ can be written explicitly as:
\eqn\klm{\eqalign{\pb \Phi_{z} + [ A_{\bar z} , \Phi_{z} ] =
\left( IJ - \zeta_{c} \Id \right) \delta^{(2)} ( z ) \cr
F_{z\zb} + [\Phi_{z} , \Phi_{\zb} ] =
\left( I I^{\dagger} - J^{\dagger} J  - \zeta_{r} \Id
\right) \delta^{(2)} ( z ) \cr}}

The solutions to \klm\ are the gauge fields which have monodromy
around the puncture $z = 0$, which is conjugate to
$\exp \left( I I^{\dagger} - J^{\dagger} J  - \zeta_{r} \Id
\right)$ and the covariantly constant Higgs fields which have a first order
 pole  at $z=0$ with the residue given by $\left( IJ - \zeta_{c} \Id \right)$.
\newsec{Beauville - Mukai's systems}

Let $S$ be the surface of $K3$ type, which contains
a holomorphically embedded curve $\Sigma$ of genus $g$. Fix the numbers
$N$ and $p$. Let $\CM_{N,p,g} (S, \Sigma)$ be
the moduli space of the pairs $(C, \CL)$, where
$[C] = N [\Sigma]$, $C$ is a
holomorphically embedded curve
and $\CL$ is the line bundle on $C$ of degree $p$. 
Let $h$ be the genus of generic $C$. It is equal to
\eqn\gns{h = 1 + N^{2}(g-1)}
Let ${\MM}$
be the compactification of $\CM_{N, h, g}(S, {\Sigma})$ (the case of
general $p$ is left beyond the scope of our investigation) which is defined as 
the moduli space 
of pairs $(C, \CL)$ where $C$ is a curve as above and
${\CL}$ is a torsion free rank one coherent 
sheaf on $C$ (if the curve $C$ is smooth
then this is the same as a line bundle), with $\int_{C} c_{1}({\CL}) = h$.

\noindent
{\bf Theorem.} $\MM$ is a symplectic manifold,
birationally equivalent to $S^{[h]}$ with its standard
symplectic structure.

\noindent
{\bf Proof.} 
The statement is actually well-known and is a compilation of two
Beauville remarks \Beaui, \Beauii. Similar considerations are contained in
\BVS,\yaz.
Fix a smooth irreducible representative $C$.
A.~Beauville proves that under the conditions of our theorem there exists a
morphism $\varphi : S \to \IP^{h}$, such that its restriction to $C$
coincides with the canonical embedding of $C$ into $\IP^{h -1}$ by the
sections of the canonical sheaf $\omega_{C}$ of $C$:  A hyperplane $H \subset
\IP^{h}$ intersects $\varphi(S)$ by a curve $C_{H}$. A restriction of the
line bundle $\CO(1)$ over $\IP^{h}$
to $C_{H}$ gives a line bundle $\CL_{H}$
of degree $h$.

Given a very ample line bundle $\CL$ over $C_{H}$
of the same  degree $h$ one may consider the corresponding linear
system $\vert \CL \vert =
\left( {\IP}^{h}\right)^{*}$ 
as a base of the fibration $\CJ \to \vert \CL \vert$
with the fiber ${\rm  Jac} (C_{H})$ \Beaui\ over a point $C_{H}$.
Moreover, as it was shown in \AltKl\ one can extend  this fibration
to a ``compactified fibration''
 $\bar\CJ$
with the fibre $\overline{{\rm Jac}(C_{H})}$
such that ${\rm dim} \bar\CJ = 2h$.
  This space is an algebraically completely integrable system
 with respect to a holomorphic symplectic
structure introduced in \Mu.
Through $h$ lineary independent points $p_1, \ldots ,p_h$ in 
$S\subset \IP^{h}$ the
unique  hyperplane $H$ passes. Hence (generically) there exists  
a unique genus $h$ curve
$C_{H}$ passing through the points and the line bundle $\CL$ whose divisor
coincides with $p_1 + \ldots + p_h$. 
This $\CL$ has a unique up to a constant non-zero section with $h$ zeroes
at the points $p_1, \ldots, p_h$.
 In other words the pairs $(C ,\CL) \in \MM$ are
parameterized by $\bar\CJ$. We have established a correspondence
between the point of $\bar\CM$  and the point of smooth open subset in
symmetric product ${\rm Sym}^{h} (S) $. Proposition $\bf 1.3$ of \Beauii\
shows that this correspondence actually extends to a
birational isomorphism of our theorem.
Thus the symplectic structure of Mukai's on $\bar\CJ$
maps to a direct sum
of symplectic structures on the copies of the surface $S$.

\noindent
{\bf Remark.} The identification of Beauville-Mukai
integrable systems phase space with
the Hilbert  scheme of points on the  initial $\bf K3$
surface $S$ provides an analogue
of Sklyanin's  $\bf SoV$ in
terms of projective coordinates in $\IP^{h}$. It again resembles 
his ``magic
recipe'':  the r\^ole of  zeroes and poles
of eigen-bundle sections is  played by the
zeroes of sections of linear bundles $\CL$ arising under replacing of
$T^{*}\Sigma$ by $\bf K3$ surface $S$.

  The following example of Beauville-Mukai system which is proposed in \Beaui\
is an illustration of this {\it ad hoc} separation.

\noindent
{\bf Example.} Let us consider a quartic $S \subset \IP^{3}$
given by the equation
 $F(X_0, X_1, X_2, X_3) = 0$ in
homogeneous coordinates $(X_0 : X_1 : X_2 : X_3)$.
This is an example of $\bf K3$ surface. The condition ${\rm deg}F = 4$
implies  that $S$ has a holomorphic symplectic form, whose associated
Poisson bracket can be described quite explicitly. Suppose $X_{0}=t \neq 0$.
Let $f(x_{1}, x_{2}, x_{3}) = t^{-4} F (t, tx_{1}, tx_{2}, tx_{3})$.
Let $g,h$ be the locally defined holomorphic functions on $S \backslash S
\cap \lbrace X_{0} = 0 \rbrace$. Extend
them to the functions in $(x_{1}, x_{2}, x_{3})$ defined in the {\neib}
of $f^{-1}(0)$. Then:
\eqn\psbr{\lbrace g, h \rbrace \equiv {{dg \wedge dh \wedge df}\over{dx_{1}
\wedge dx_{2} \wedge dx_{3}}}}
evaluated at $f =0$
is well-defined independently of the choices made and also the Poisson
bivector defined in this way extends to the whole of $S$.
Now we want to study the $h=3$'d symmetric power of $S$, more precisely
the Hilbert scheme $S^{[3]}$.
Introduce the variables $X_{ai}$, $i=1, \ldots, 3$,
$\alpha = 0, 1,2,3$. We denote by $\vec x_{i} =
\left( x_{1i}, x_{2i}, x_{3i} \right)$, where $x_{a i } = X_{a i }/X_{0 i}$, $a 
= 1,2,3$.
The space $S^{[3]}$ has a
natural sympectic form $\omega_{h}$ which is the induced from
the symplectic form on $S$ via the Hilbert-Chow morphism
$S^{[3]} \to {\rm Sym}^{3}(S)$. Let $\pi = \omega_{h}^{-1}$ be 
the corresponding Poisson
bivector. 
We now define three functions on $S^{[3]}$. Assume first that $
\vec x_{i} \neq \vec x_{j}, i \neq j$.  Then the construction of the
previous section can be formulated very concretely as follows. There is a
unique hyperplane in $\IP^{3}$ which passes through the points
$\vec x_{1}, \vec x_{2}, \vec x_{3} \in S \subset \IP^{3}$. The space of
hyperplanes in $\IP^{3}$
is the dual projective space $\CH = \left( \IP^{3} \right)^{\star}$:
\eqn\dlprsp{c = (c^{0}: c^{1}: c^{2}: c^{3}) \in \CH \mapsto
\sum_{a=0}^{3} c^{a} X_{a} = 0, \quad X = (X_{0} : X_{1}: X_{2} :X_{3}) \in
\IP^{3}}

The Hilbert scheme of points contains regions where the points coincide.
For a collision of a pair of points $\vec x_{1}, \vec x_{2}$
one glues a line $\IP^{1}$ which
determines
the direction along which the points collided. In this case
there still exists a unique hyperplane in $\IP^{3}$ which passes through
this line and the third point $\vec x_{3}$. Now, if the third point
approaches the cluster formed by the first two then the plane $\IP^2$
along which
the configuration of the line and the third point collided is contained in
the Hilbert scheme of points. Thus we have shown that by associating a
plane to the triple of points and they limits one gets a well-defined
birational map:
\eqn\mkh{\varphi: S^{[3]} \to \CH}
The map
$\varphi$ is given by the explicit formulae:
\eqn\mkhm{{\varphi}^{\alpha} = {\Det} \Psi_{\alpha}}
where $\Psi_{\alpha}$ is the $3 \times 3$ matrix obtained from $\Vert X_{ai} 
\Vert$ by removing
the $\alpha$'th  column. In the domain where $\psi_{0} := {\rm Det} \Vert x_{ai} 
\Vert \equiv
\Psi_{0} / \left( X_{10}X_{20}X_{30} \right) \neq 0$ we may introduce the 
functions:
\eqn\hmltn{H_{a}  =
{{\psi_{a}}\over{\psi_{0}}}}
where $\psi_{a} = {\rm Det}M_{a}$, $\left( M_{a} \right)_{bi} = X_{bi}$, for $b 
\neq a$ and
$\left( M_{a} \right)_{ai} = 1$ for any $i$:
\eqn\hmltns{\eqalign{&
H_1 = {1\over{\psi_0}} {\Det} \pmatrix{1& x_{21}&x_{31}\cr
         1& x_{22}&x_{32}\cr
         1& x_{23}&x_{33}\cr}, \cr
& H_2 = {1\over{\psi_0}} {\Det} \pmatrix{x_{11}& 1&x_{31}\cr
            x_{12}&1& x_{32}\cr
            x_{13}&1& x_{33}\cr}, \cr
& H_3 = {1\over{\psi_0}} {\Det} \pmatrix{x_{11}&x_{21}&1\cr
            x_{12}&x_{22}&1\cr
            x_{13}&x_{23}&1\cr},\cr}}
These explicit Hamiltonians
are defined in $\varphi^{-1}(\IA^{3})$, where $\IA^{3}$ is the
affine part of
$\CH$, corresponding to $\psi_{0} \neq 0$.
It follows from the  identity:
\eqn\indties{\{ \psi_{a}, \psi_{b} \} = \psi_{a} \{ \psi_{0}, \psi_{b} \} - 
\psi_{b} \{ 
\psi_{0}, \psi_{a} \} }
that the Hamiltonians $H_{a}$, $a = 1,2,3$, Poisson commute.
It would be interesting to investigate this system as an example
of a ``deformation'' of a Hitchin system in the sense of
\DonEinLaz\ and to study its quantum analogues which as are hopefully 
related to some Feigin - Odesski - Sklyanin algebras.

\newsec{Relations to the physics of $D$-branes and gauge theories}

The abovementioned constructions of the separation of variables
in integrable systems on moduli spaces of holomorphic bundles
with some additional structures can be described
as a symplectomorphism between  the moduli spaces of
the bundles (more precisely, torsion free sheaves) having different topology,
e.g. Chern classes.

To be specific let us concentrate on the moduli space
$\CM_{\vec v}$ of stable torsion free coherent sheaves ${\CE}$
on $S$. Let ${\hat A}_{S} = 1 - [ {\rm pt} ] \in H^{*} (S, {\IZ})$ 
be the $A$-roof
genus of $S$. The vector
$\vec v = Ch ({\CE}) \sqrt{\hat A_{S}} =
(r; \vec w; d - r) \in {\H}^{*}(S, {\IZ}), \vec w \in \Gamma^{3,19}$
corresponds to the sheaves
 with the Chern numbers:
\eqn\mkvctr{\eqalign{
ch_{0} ({\CE}) & =
 r \in {\H}^{0}(S ; {\IZ})\cr
ch_{1}({\CE}) & = \vec w \in {\H}^{2} (S; {\IZ}) \cr
ch_{2}({\CE}) & = d \in {\H}^{4}(S; {\IZ}) \cr}}
Type $\II A$ string theory compactified on $S$ has BPS
 states, corresponding to
the $Dp$-branes, with $p$ even, wrapping
various supersymmetric cycles in $S$, labelled by
$\vec v \in {\H}^{*}(S, {\IZ})$. The actual states
correspond to the cohomology classes of the moduli spaces
$\CM_{\vec v}$
of the configurations of branes.  The latter can be identified
with the moduli spaces $\CM_{\vec v}$ of appropriate sheaves.

The string theory, compactified on $S$ has moduli space
of vacua, which can be identified with
$$
\IM_{A} = O\left( {\Gamma}^{4,20} \right) \backslash
O(4,20; {\IR}) / O(4;{\IR}) \times O(20;{\IR})
$$
where the arithmetic group $O({\Gamma}^{4,20})$ is the group
of discrete authomorphismes. It maps the states
corresponding to different $\vec v$ to each other.
The only invariant of its action is ${\vec v}^{2}$.
We have studied three realizations of an
integrable system. 

The first one uses the non-abelian
gauge fields on the curve $\Sigma$ imbedded into symplectic
surface $S$. Namely,  the phase space of the system is the
moduli space of stable pairs: $(\CE, \phi)$, where $\CE$
is rank $r$ vector bundle over $\Sigma$ of degree
$l$, while $\phi$
is the holomorphic section of
$\omega^{1}_{\Sigma} \otimes {\rm End}({\CE})$.

The second realization is the moduli space of pairs $(C, {\CL})$,
where $C$ is the curve (divisor) in $S$ which realizes
the homology class $r[\Sigma]$ and $\CL$ is the line bundle on $C$.

The third realization is the Hilbert scheme of points on $S$
of length $h$, where $h = {\half}{\rm dim}{\CM}$.

The equivalence of the first and the second realizations
corresponds to the physical statement that the bound
states of $N$ $D2$-branes wrapped around $\Sigma$ are represented
by a single $D2$-brane which wraps a holomorphic curve $C$
which
is an $N$-sheeted covering of the base curve $\Sigma$. 
The equivalence of the second and the third descriptions
is tempting to attribute to $T$-duality.

\newsec{Discussion}

We have attempted to formulate the separation of variables
purely in geometric terms. It seems that the proper physical setup
is the use of chain of dualities to get the system of $D0$ branes
on some hyperkahler  manifold\foot{As our manuscript was at the stage of final
preparation, a very interesting paper \dijkgraaf\ appeared which studied
various aspects of string theory on instanton moduli spaces on 
hyperkahler manifolds. It also contains an extensive
discussion of the duality groups}.

A few subjects need  further clarification. Evidently an 
interesting question to investigate is the quantization
of the integrable system. According  to the work of E.~Frenkel and
B.~Feigin,
the
quantum separation of
variables can be viewed as the geometrical Langlands transform.
The latter
maps the spectrum of the Hitchin $\CD$-module (Hamiltonians)
to the data connected to the local system on the dual object \beil.
What we have studied  is the classical analogue of the Langlands
correspondence which maps the phase space of the Hitchin and
Mukai systems to the Hilbert scheme of points.

The quantum  separation of variables goes
as follows. The wave function in the separated variables becomes the
product of identical factors  up to a determinant factor
$\prod_{i,j} (\xi_i - \xi_j)^{\kappa}$:
$$
\Psi(\xi_{1},\ldots ,\xi_{n})= \prod_{i} Q( \xi_{i})
$$
where $Q(\xi )$ obeys the so-called Baxter's equation
$$
{\Det} \left( L(\xi) -\eta\right) Q(\xi)=0
$$
where $L(\xi)$ is the Lax operator of the dynamical system,
evaluated at the zero of the Baker-Akhiezer function.
The variable $\eta$ is viewed as an
operator acting on $Q(\xi)$.
The commutation relation between $\xi$ and $\eta$
implies the representation of $\eta$, which can be either
differential, difference or
integral operator.

In some sense  $Q(\xi_{i})$ can be considered as the
wave function of a single $D0$ brane while the union of integrable
systems with all $N$'s is the second quantization.

As an example of the quantum separation of variables
let us consider the Gaudin system.
Consider $\CD$-modules  which
describe a system of differential equations on the moduli space
${\CM}_{G} (\Sigma )$ of principal $G$-bundles over a complex curve
$\Sigma$
with marked points which is given by a commuting set of differential
operators whose symbols coincide with Hitchin hamiltonians.
Hence they can be considered as a quantization of Hitchin systems \beil.
If the curve $\Sigma$ has genus zero then the  corresponding $\CD$-modules
give rise to
the Gaudin model of integrable quantum spin chain \FFR.

Consider the punctured sphere $\Sigma = \IC\IP^1$
and attach to each of the distinct marked
points $z_{i} , i=1,...n$ the $\lieg =sl_2$
Verma module $V_{\lambda_i}$ of the highest
weight $\lambda_i \in \IC$.
Represent the  Lie algebra $sl_2$  by
the  operators
\eqn\slt{
e_i = - t_i{\p}^{2}_{t_i} -  {\lambda_i}{\p}_{t_i},
\quad h_i =- 2t_i{\p_{t_i}} - {\lambda_i},
\quad f_i = {t_i},}
and  denote by $C_{ij}$ the Casimir operator
$C_{ij} = e_{i}f_{j} + f_{i}e_{j} + {\half} h_{i}h_{j}$.
The Lax operator
$\widehat{\phi(z)}$ is given by
\eqn\lxopr{\widehat{\phi(z)} = \sum_{i=1}^{n}
{{\pmatrix{h_{i} &  f_{i} \cr e_{i} & -h_{i}
\cr}}\over{z - z_{i}}} }
Introduce the operators $H_i$ :
\eqn\oph{
H_i = \sum_{i {\neq} j} {{C_{ij}}\over{z_i - z_j}},
}
and $\nabla_{i} = (\kappa + 2)\p_{z_i} - H_i$, where $\kappa \in \IC$
is a ``level''.
Consider the infinite-dimensional Lie algebras $\hat\lieg$ and $\hat\lieg_{+}$:
$\hat\lieg = \lieg \otimes \IC ((z)) \oplus \IC K$,
where $\IC ((z))$ are meromorphic
functions of a formal parameter $z$ around a marked
 point and
$\hat\lieg_{+} = \lieg\otimes\IC [[z]]$ is the Lie subalgebra of $\hat\lieg$
which consists of the  power seria in $z$. The quotient
$U\hat\lieg / (K-\kappa)$ is
denoted by $U_{\kappa}\hat\lieg$.
By definition the conformal blocks are  linear
functionals:
\eqn\cnfrbl{f :
\otimes_{i=1}^{n}(U_{\kappa}(\hat\lieg) \otimes_{U(\hat\lieg_{+})}
V_{\lambda_i}) \to \IC,}
which are
invariant under the action of the Lie algebra $\hat\lieg^{\rm out}
(z_1,...,z_n)$ of
the Laurent seria at the marked points of regular $\lieg$-valued functions on
the
marked sphere which vanish at infinity.
It is well-known fact \FFR\ that any such  functional defines  a
number-valued function $f(z_1,\ldots ,z_n; t_1,\ldots, t_n)$ 
obeying the
Knizhnik-Zamolodchikov system
\eqn\kz{\nabla_i (f) = 0.}
Let us perform the transition to the separated variables $p_{i}$: 
$$
(z, t_1, \ldots, t_n) \to (z,p_1, \ldots, p_{n-1},u),
$$
where the variables $(p, u)$ and $t$ are defined through the
following relation:
\eqn\rela{
\sum_{i=1}^{n} {{t_i}\over{z-z_i}} \,dz = u
{{\prod_{l=1}^{n-1} (z-p_l)}\over{\prod_{i=1}^{n}(z-z_i)}}\,dz. }
It is clear that $p_{l}$ which are the
zeroes of the $\Phi_{12}$ element of the
Lax matrix are zeroes of the Baker-Akhiezer function.
Therefore  the spectral curve
$$
\eta^{2}   =  {\Det}\Phi(z)
$$
yields Baxter-Sklyanin equation for the Gaudin magnet
$$
\left[ \nabla^{2}_{p_{i}}  
+ {\Det} {\Phi} (p_{i}) \right] Q(p_{i}) =0, 
$$
where $\nabla = -\p_{p} - \sum_{i} {{\lambda_i}\over{p - z_i}}$, 
which defines a projective connection. If one 
considers $\lieg = {\rm sl}_{N}$ magnet instead then
Baxter equation becomes the $N$-th order differential equation.

In generic situation there are a few subtle points.
First, the proper ordering
should be chosen in the quantum operators. Secondly, when discussing
Hitchin system on higher genus curve one has to deal with globally defined
objects with proper modular properties.
There are no global solutions to the differential equations in this
case. The differential operators are defined globally, though.

The
explicit computations in the case of genus one \er\nhol\ give a system
which interpolates between the Gaudin model and Calogero-Moser elliptic
many-body system. The geometrical approach to {\bf SoV} 
for this system is formulated in \Fr\ and elaborated on in
\efr.

\centerline{\bf Acknowledgements}
We would like to thank
B.~Feigin, E.~Frenkel, V.~Ginzburg, B.~Enriquez, L.~Evain,
I.~Krichever, S.~Kharchev, A.~Losev, I.~Reider,  
A.~Stoyanovsky and especially D.~Markushevich
for useful and interesting discussions and to
V.~Fock for collaboration at the initial stage of the project.

Research of A.~G. is supported by grants CRDF-RP2-132, 
INTAS 96-482 and {\cyr RFFI} 97-02-16131; 
that of  N.~N.~ is supported by Harvard Society of Fellows, 
partially by NSF under
grant PHY-98-02-709, partially by {\cyr RFFI}
 under grant 96-02-18046 and partially
by grant 96-15-96455 for scientific schools. He is
  also grateful to Aspen Center for Physics
for hospitality while part of this work was done. 
V.~R.~ is partly supported by the grant INTAS-96-196, by 
{\cyr RFFI} under the grants
98-01-00327, by the grants 
for promotion of French-Russian scientific cooperation: 
CNRS grant PICS No. 608, {\cyr RFFI}  98-01-2033, 
 and by the grant 96-15-96455 for support of scientific schools. 
V.~R.~ is also 
grateful to Universit\'e Louis Pasteur in Strasbourg for
its kind hospitality  during the Spring term of 1997. His particular thanks
are due to M.~Audin, O.~Debarre and J.-I.~Merindol for numerous discussions.

We would like to thank Mittag-Leffler Institute for hospitality 
at the final stage of preparation of the manuscript.

\listrefs
\bye